\documentclass{aa}
\usepackage{times}
\usepackage{graphics}
\begin{document}

\thesaurus{08(03.19.2; 08.09.2 BD +39 3226; 09.01.1; 09.13.2; 12.03.2; 13.21.3)}

\title{{\sl ORFEUS II} echelle spectra:\\
deuterium and molecular hydrogen in the ISM towards BD\,+39\,3226\thanks{Based on data obtained under the {\sl DARA ORFEUS} guest observer programme}}

\author{H. Bluhm \inst{1}
        \and
        O. Marggraf \inst{1} 
        \and
        K.S. de Boer \inst{1}
        \and
        P. Richter \inst{1}
        \and
        U. Heber \inst{2}} 

\offprints{hbluhm@astro.uni-bonn.de}

\institute{
Sternwarte der Universit\"at Bonn, Auf dem H\"ugel 71, D-53121 Bonn, Germany
\and
Dr. Remeis-Sternwarte, Astronomisches Institut der Universit\"at Erlangen-N\"urnberg, Sternwartstrasse 7, D-96049 Bamberg, Germany
}

\date{Received / Accepted}

\titlerunning{Deuterium and molecular hydrogen in the ISM towards BD\,+39\,3226}
\authorrunning{H. Bluhm et al.}
\maketitle

\begin{abstract}
 In {\sl ORFEUS II} spectra of the sdO star \object{BD +39\,3226} interstellar hydrogen and deuterium is detected.
 From Ly $\alpha$  profile fitting and a curve of growth analysis of the Lyman~series of \ion{H}{i} and \ion{D}{i} we derive the column densities $N_{\rm H} = 1.20^{+0.28}_{-0.22}{\cdot}10^{20}$\,cm$^{-2}$ and $N_{\rm D} = 1.45^{+0.50}_{-0.38}{\cdot}10^{15}$\,cm$^{-2}$.
 From the analysis of metal absorption lines in {\sl ORFEUS} and {\sl IUE} spectra we obtain column densities for 11 elements.     
 In addition, we examine absorption lines of H$_2$ for rotational excitation states up to $J=7$. 
 We find an H$_2$ ortho-to-para ratio of 2.5, 
 the fractional abundance of molecular hydrogen has a low value of $\log f=-4.08$ for a total amount of $N(\mbox{H}_2)=4.8^{+2.0}_{-1.6}{\cdot}10^{15}$cm$^{-2}$.
 The column densities of the excitation states reveal a moderate Boltzmann excitation temperature of $130$~K and an equivalent excitation temperature for the excited upper states due to UV pumping of $<1800$~K.

 
\end{abstract}

\section{Introduction}
The deuterium abundance in the Galaxy has been of interest since the first models of primordial nucleosynthesis were developed.
 While it is unlikely that major amounts of deuterium could be produced after the Big Bang, it is sure that deuterium is almost completely destroyed in stars.
 This should lead to a decrease of its abundance in the progress of galactic evolution.
 Every measurement of $D/H$ thus should give a lower limit for the primordial abundance and, according to nucleosynthesis models, an upper limit for the baryonic density in the universe.
 Today, these limits are set by measurements in quasar spectra.
 For example Levshakov et al. (\cite{lev}) suggest a uniform $D/H \simeq 4{\cdot}10^{-5}$ for three lines of sight at redshifts of $z$ between  $0.7$ and $3.6$, but Molaro et al. (\cite{mola}) find $1.5\cdot10^{-5}$ from a line of sight with $z=3.5$. 
 Nevertheless knowledge of abundances in the galactic ISM can be important for tracing the evolution of the Milky Way. 

Since the {\sl Copernicus} satellite in the 1970s no instrument was capable of high resolution spectroscopy in the $900$ to $1200$~\AA\ range.
 There have been $D/H$ measurements with the {\sl IUE} and the {\sl HST-GHRS}, but only the Ly\,$\alpha$ line was observable leading to a restriction to lines of sight with low hydrogen column densities.
 For an overview see e.g. Lemoine et al. (\cite{lemo}).

The {\sl ASTRO-SPAS} space shuttle platform housed 3 spectrographs operating between $900$ and $1200$ {\AA}.
Of those, {\sl IMAPS} has an echelle spectrograph working between $930$ and $1150$~{\AA} with a resolution of $\lambda/{\Delta\lambda}\simeq 1.5\cdot10^5$.
 Jenkins et al. (\cite{jenk2}) and Sonneborn et al. (\cite{sonne}) performed $D/H$ measurements with that instrument.   
 Two spectrographs were attached to the {\sl ORFEUS} telescope.
 The Berkeley spectrograph is designed for intermediate-resolution spectra in the range of $390$ to $1220$~{\AA}.  
 The Heidelberg-T\"ubingen~ echelle spectrograph, equipped with a microchannel plate detector, gives spectra from $912$ to $1410$~{\AA} and allows investigations of the entire hydrogen and deuterium Lyman series with a resolution of $\lambda/{\Delta\lambda}\,{\simeq}\,10^4$ (Kr\"amer et al. \cite{kraem}).
 We use spectra of the latter to investigate deuterium.

 The large number of rotational transitions of molecular hydrogen found in the wavelength range from 1200~\AA\, up to the Lyman edge can be used for studies on the molecular gas along the line of sight.
 The distribution of molecular hydrogen in the local interstellar medium 
especially at higher galactic latitudes is still only rudimentarily known,
since the majority of the {\sl Copernicus} measurements pointed towards 
luminous targets in the galactic plane.
From the H$_2$ transitions column densities of the different H$_2$ rotational 
excitation states up to $J=7$ can be determined.
These can be used to derive physical parameters as the excitation temperature 
and the ortho-to-para ratio of the molecular hydrogen.

\section{Observations and data reduction}

 \object{BD +39\,3226} is a sdO star  displaying a pure helium line spectrum in the optical (Giddings \cite{gidd}) in a distance of about 270~pc\footnote{Giddings (\cite{gidd}) found a spectroscopical distance of $240^{+90}_{-70}$\,pc, the {\sl Hipparcos} distance is $290^{+140}_{-70}$\,pc} at the galactic coordinates $l = 65.00$, $b = +28.77$. 
Its high radial velocity of $\simeq -$279~km\,s$^{-1}$ makes this star suitable
for studies of the local ISM, since stellar and interstellar absorption
lines are well separated.
 We analyse echelle spectra obtained during the {\sl ORFEUS II} 
mission in Nov./Dec. 1996 and {\sl NEWSIPS} reduced long- and short-wavelength high dispersion spectra taken from the {\sl IUE Final Archive} 
({\sl IUEFA})\footnote{WWW URL: {\tt http://iuearc.vilspa.esa.es}}.
 The total wavelength range thus covered is about 900 to 3200~\AA.

 The {\sl ORFEUS} spectrum was obtained in 4 pointings with a total of $5000$~s integration time.
 The basic reduction of the data was performed by the {\sl ORFEUS} team in T\"ubingen.
 Details about the {\sl ORFEUS} instrumentation and the basic data handling and calibration are given in Barnstedt et al. (\cite{barn}).
 We repeat the main features here.

 The wavelength calibration of the {\sl ORFEUS} spectra is based on the interstellar spectrum in the {\sl ORFEUS} target \object{HD 93521}.
 The accuracy of this calibration is estimated as better than $0.05$~{\AA}, 
 but small systematic effects within the spectrum cannot be excluded.
 The zero point of the wavelength calibration is based on the  geocoronal Ly $\alpha$ emission.
 Since the {\sl ORFEUS} aperture measures 20$^{\prime\prime}$ in diameter, errors in the pointing may shift the wavelength zero point.
 Optical spectra of BD~+39\,3226 give the stellar heliocentric radial velocity as $-278.7\pm5$~km\,s$^{-1}$, based on values by Dworetsky et al. (\cite{dwor}) and Giddings (\cite{gidd}) and one derived from observations with {\sl FOCES} at the Calar Alto $2.2$~m telescope.
 In the {\sl ORFEUS} spectrum the mean radial velocity of $22$ sharp stellar metal lines is $-289.1\pm1.5$~km\,s$^{-1}$.
 We corrected the {\sl ORFEUS} spectrum for this shift.  
 The new velocity zero point should be accurate within $\approx5$~km\,s$^{-1}$.

  The spectral resolution is intrinsically better than 10$^4$, 
but pointing jitter ($\simeq 2^{\prime\prime}$) 
and the addition of spectra from several pointings 
may cause some deterioration. 
We have filtered the spectra as provided by T\"ubingen 
with a wavelet algorithm (Fligge \& Solanki 1997) 
effectively leading to a 3 pixel boxcar filter. 
We derive from our spectrum a resolution of 
$\lambda/\Delta\lambda \simeq 0.9{\cdot}10^4$. 

 {\sl ORFEUS} spectra are influenced by scattered light, 
which is implicitly corrected for by subtracting 
the intensities in the interorder space from 
the echelle order intensities. 
 Near strong features residual effects may still be present (Barnstedt et al. \cite{barn}).
 However, since the {\sl ORFEUS} spectra are slightly tilted 
with respect to the detector grid, 
affected areas lie normally at some distance from such features.

 The resolution of the {\sl IUE} spectra is also about $10^{4}$, which is equivalent to ${\simeq} 30~$km\,s${^{-1}}$.
 While only one long wavelength spectrum  (LWR 11789, exp. time 5040\,s) can be found in the {\sl IUEFA}, several short wavelength spectra are available (SWP 15275, 48312, 48313, 48314, each with 3600 s exp. time).  
 To improve  the S/N ratio we added the four SWP spectra.
 The {\sl IUE} spectra were smoothed with a 3 pixel boxcar filter.

\section{Column densities}

\subsection{Data analysis}

 We identified numerous interstellar absorption lines of H\,{\sc i}, D\,{\sc i}, different heavy elements and H$_2$.
  In the photospheric spectrum only a few weak, sharp metal lines of C\,{\sc iii}, C\,{\sc iv}, N\,{\sc iii}, N\,{\sc v}, Si\,{\sc iii}, Si\,{\sc iv}, P\,{\sc v}, and S\,{\sc v} can be identified besides the strong He\,{\sc ii} lines.
 The investigations on interstellar metal and deuterium abundances make use of the 
standard curve of growth technique. 
 Except for the case of the Ly ${\alpha}$ and ${\delta}$ absorptions, the equivalent widths of the lines were measured either by a trapezium or by a Gaussian fit.
 The differences between the two methods are well below the typical errors in 
the equivalent widths. 
 Uncertainties in the measurements occur because of noise and the choice of the continuum. 
 The error due to the latter was estimated by determining the equivalent width for a lower and an upper limit of the continuum level, the photon statistics were taken into consideration by a formula adopted from Jenkins et al. (\cite{jenk}). 
  Both errors added quadratically give the uncertainties in $W_{\lambda}$ as quoted in Tables \ref{eqw1}, \ref{eqw2}, and \ref{H2_ew}.  
 
 Some lines of S\,{\sc ii}, Si\,{\sc ii}, and N\,{\sc i}, which lie between 1190 and 1390 \AA,
were measured in both the {\sl IUE} and the {\sl ORFEUS} spectrum.
 The line profiles and absorption strengths turn out to be consistent.  

\subsection{Metals}

\begin{figure}[t!]
\resizebox{\hsize}{!}{\includegraphics{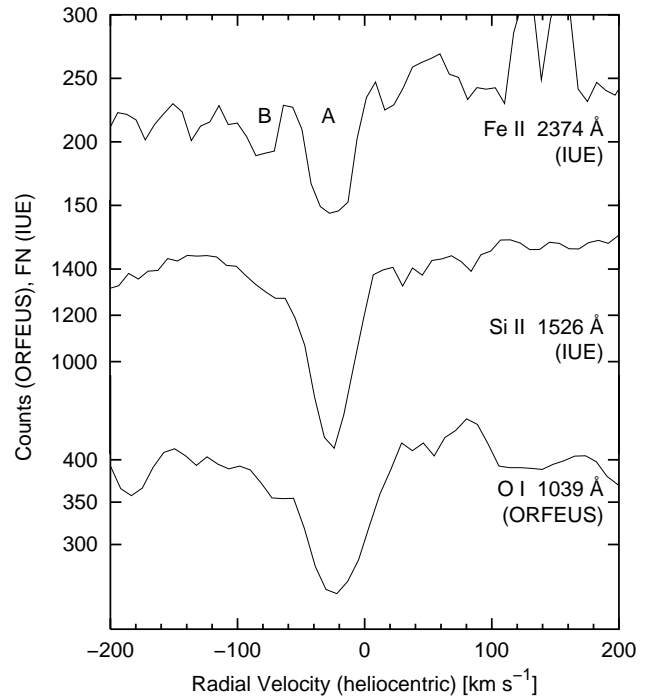}}
 \caption{Profiles of metal absorption lines in the {\sl IUE} LWR, {\sl IUE} SWP, and the {\sl ORFEUS} spectrum. The velocity components (heliocentric) at $v_{\rm rad} \simeq -75$~km\,s$^{-1}$ (B) and $v_{\rm rad} \simeq -24$~km\,s$^{-1}$ (A) are labeled} 
\label{velstruc}
\end{figure}

\begin{table}[t!]
\begin{minipage}[t]{9cm}
\caption[]{Metal absorption line equivalent widths measured in component A (at~ $v_{\rm rad} \simeq -24$~km\,s$^{-1}$) with {\sl ORFEUS} and {\sl IUE}
  }
\label{eqw1}
\setlength{\tabcolsep}{1.65mm}
\begin{tabular}{lllcc}
\hline\noalign{\smallskip}
Ion & Wavelength & $\log(f\lambda)$ &  W${_\lambda}$ (\sl ORFEUS)& W${_\lambda}$ (\sl IUE) \\ 
  & \multicolumn{1}{c}{[\AA]} &  &[m{\AA}] & [m{\AA}] \\ \noalign{\smallskip}\hline\noalign{\smallskip}  
Fe\,{\sc ii}& $1096.8769$ & $1.545$ & $25.6\pm\hspace{0.17cm}8.0$ & -- \\
       & $1125.4477$ & $1.092$ & $12.8\pm\hspace{0.17cm}6.4$ & -- \\
       & $1144.9379$ & $2.080$ & $64.1\pm\hspace{0.17cm}6.9$ & -- \\
       & $1608.4511$ & $1.998^*$ & -- & $112.1\pm18.8$\hspace{0.17cm} \\
       & $2344.214$ & $2.410$ & -- & $161.9\pm16.4$\hspace{0.17cm} \\
       & $2374.4612$ & $1.889^*$ & -- & $101.4\pm15.8$\hspace{0.17cm} \\
       & $2382.039$ & $2.967$ & -- & $163.9\pm15.9$\hspace{0.17cm} \\
       & $2586.6500$ & $2.248^*$ & -- & $161.3\pm26.7$\hspace{0.17cm} \\
       & $2600.1729$ & $2.765$ & -- & $189.1\pm20.8$\hspace{0.17cm} \\ 
Si\,{\sc ii}& \hspace{0.17cm}$989.8731$ & $2.119$ & $66.0\pm15.5$&   --  \\
       & $1190.4158$ & $2.474$ & $80.2\pm11.1$ & $79.4\pm\hspace{0.17cm}7.9$ \\
       & $1193.2897$ & $2.775$ & $101.1\pm10.0$\hspace{0.17cm} & $94.5\pm\hspace{0.17cm}7.1$ \\
       & $1260.4221^a$& $3.104$ & $98.8\pm16.4$  & $92.9\pm11.4$  \\
       & $1304.3702$ & $2.050{^*}$ & $69.6\pm15.2$ & $70.9\pm\hspace{0.17cm}6.5$ \\ 
       & $1526.7066$ & $2.225{^*}$ & -- & $105.4\pm\hspace{0.17cm}8.1$\hspace{0.17cm} \\
       & $1808.0126$ & $0.596{^*}$ & -- & $33.3\pm\hspace{0.17cm}8.4$ \\
Si\,{\sc iii}& $1206.500$ & $3.311$ & $69.6\pm10.9$ & $64.0\pm12.9$ \\
N\,{\sc i}& \hspace{0.17cm}$964.6256$ & $1.059$ & $31.2\pm\hspace{0.17cm}8.7$ & -- \\
     & $1134.9803$ & $1.660$ & $68.5\pm10.2$ & -- \\
     & $1199.5496$ & $2.202$ & $115.2\pm10.4$\hspace{0.17cm} & $80.0\pm\hspace{0.17cm}5.2$ \\
     & $1200.2233$ & $2.026$ & $82.1\pm\hspace{0.17cm}7.6$ & $62.5\pm\hspace{0.17cm}7.8$ \\
     & $1200.7098$ & $1.725$ & $54.9\pm\hspace{0.17cm}9.2$ & $70.5\pm\hspace{0.17cm}9.6$ \\
O\,{\sc i}& \hspace{0.17cm}$924.952$ & $0.168$ & $49.4\pm16.7$ & -- \\
     & \hspace{0.17cm}$936.6295$ & $0.544$ & $67.2\pm11.9$ & -- \\
     & \hspace{0.17cm}$950.8846$ & $0.174$ & $57.7\pm\hspace{0.17cm}7.0$ & -- \\
     & \hspace{0.17cm}$976.4481$ & $0.508$ & $65.1\pm11.1$ & -- \\
     & $1039.2304$ & $0.980$ & $70.4\pm\hspace{0.17cm}9.7$ & -- \\
     & $1302.1685$ & $1.804$ &  $b$ & $b$ \\
C\,{\sc i}& $1277.2454$ & $2.091$ & -- & $20.5\pm\hspace{0.17cm}3.4$ \\
     & $1560.3092$ & $2.099$ & -- & $23.4\pm\hspace{0.17cm}8.8$ \\
     & $1656.9283$ & $2.367$ & -- & $45.9\pm\hspace{0.17cm}7.7$ \\
C\,{\sc ii}& $1036.3367$ & $2.124$ & $94.4\pm10.7$ & -- \\
      & $1334.5323$ & $2.235$ & $134.0\pm15.6$\hspace{0.17cm} & $142.4\pm\hspace{0.17cm}4.0$\hspace{0.17cm} \\  
S\,{\sc ii}& $1250.584$ & $0.834$ &$34.2\pm10.0$ &$36.3\pm\hspace{0.17cm}5.4$ \\
      & $1253.811$ & $1.135$ &$22.2\pm\hspace{0.17cm}7.5$ &$46.4\pm\hspace{0.17cm}6.3$ \\
      & $1259.519$ & $1.311$ &$50.0\pm10.9$ & $59.2\pm\hspace{0.17cm}8.0$ \\
Mg\,{\sc i}& $2852.9642$ & $3.718$ & -- & $79.9\pm17.1$ \\
Mg\,{\sc ii}& $2796.352$ & $3.235$ & -- & $239.7\pm14.9$\hspace{0.17cm} \\
       & $2803.531$ & $2.934$ & -- & $250.6\pm20.6$\hspace{0.17cm} \\
Mn\,{\sc ii}& $2576.877$ & $2.956$ & -- & $56.7\pm11.2$ \\
       & $2594.499$ & $2.847$ & -- & $51.9\pm13.0$ \\
Ni\,{\sc ii}& $1370.132$ & $2.254$ & -- & \hspace{0.17cm}$6.9\pm\hspace{0.17cm}5.0$ \\
       & $1741.549$ & $2.256$ & -- & $14.5\pm\hspace{0.17cm}5.8$ \\
Zn\,{\sc ii}& $2026.136$ & $2.996^*$ & -- & $43.9\pm14.2$ \\
       & $2062.664$ & $2.723^*$ & -- & $10.3\pm\hspace{0.17cm}6.9$ \\
Al\,{\sc ii}& $1670.7874$ & $3.486$ & -- & $92.0\pm\hspace{0.17cm}6.7$ \\
\noalign{\smallskip}\hline\noalign{\smallskip}         
\end{tabular}
\noindent
Wavelengths and $f$-values are from Morton (\cite{mor}), except for \hspace{0.05cm} $^*$~taken from
Savage \& Sembach (\cite{sav:sem})\\
$^a$ Line blended with \ion{Fe}{ii} at $1260.533$~\AA. The equivalent width given here has already been corrected for a contribution of $31.5\pm4.6$~m{\AA} derived from the determined \ion{Fe}{ii} column density \\
$b$ The $1302$ \ion{O}{i} line is in {\sl ORFEUS} spectra blended with geocoronal \ion{O}{i} emission, in {\sl IUE} spectra with a reseaux  
\end{minipage}
\end{table}

\begin{table}[]
\caption{Metal column densities $N$ [cm$^{-2}$] and abundances. Solar values are taken from de Boer et al. (\cite{boer})}    
\label{colden1} 
\begin{tabular}{llll}
\hline\noalign{\smallskip}
Element & $\log N$ & $\log (N/N_{\rm H})$ & $\log (N/N_{\rm H})_{\sun}$  \\ \noalign{\smallskip}\hline\noalign{\smallskip}
C  & $16.40\,\pm0.75$ & $-3.70\,\pm0.75$ & $-3.4$ \\ 
N  & $14.75\,\pm0.25$ & $-5.35\,\pm0.25$ & $-4.0$ \\
O  & $16.40\,^{+0.75}_{-0.50}$ & $-3.70\,^{+0.75}_{-0.50}$ & $-3.16$ \\
Mg & $14.95\,\pm0.80$ & $-5.15\,\pm0.80$ & $-4.46$ \\  
Al & $13.00\,^{+0.40}_{-0.35}$ & $-7.10\,^{+0.40}_{-0.35}$ & $-5.6$ \\
Si & $14.80\,\pm0.20$ & $-5.30\,\pm0.20$ & $-4.45$ \\
S  & $14.80\,\pm0.20$ & $-5.30\,\pm0.20$ & $-4.8$ \\
Mn & $12.60\,\pm0.10$ & $-7.50\,\pm0.15$ & $-6.58$ \\
Fe & $14.10\,\pm0.15$ & $-6.00\,\pm0.15$ & $-4.50$ \\
Ni & $12.65\,\pm0.30$ & $-7.45\,\pm0.30$ & $-5.72$ \\
Zn & $12.40\,\pm0.30$ & $-7.70\,\pm0.30$ & $-7.4$ \\
\noalign{\smallskip}\hline  
\end{tabular} 
\end{table}

In order to judge the velocity structure of the absorption it is reasonable to analyze the metal lines at first.  
We find 2 different absorption components: one at a radial velocity (heliocentric) of ${\simeq-24}$~km\,s$^{-1}$  (component A) and one at $\simeq-75$~km\,s${^{-1}}$ (component B).
 The latter is rather weak,
 Si\,{\sc ii} equivalent widths are smaller than 24 m{\AA} with large uncertainties,
 so an examination of component B can give only very uncertain results.
 Even if this component had a very low $b$-value, the upper limits for the \ion{S}{ii} and \ion{Si}{ii} column densities would be roughly $1.6{\cdot}10^{14}$\,cm$^{-2}$ and $7.9{\cdot}10^{14}$\,cm$^{-2}$ respectively corresponding to a hydrogen column density of $\leq 10^{19}$\,cm$^{-2}$ .

 In the following we will concentrate on the cloud at $-24$~km\,s${^{-1}}$. The small LSR velocity of this component suggests a local origin while component B at $-75$~km\,s$^{-1}$ represents most likely a cloud at larger distance.

\begin{figure*}
\resizebox{12cm}{!}{\includegraphics{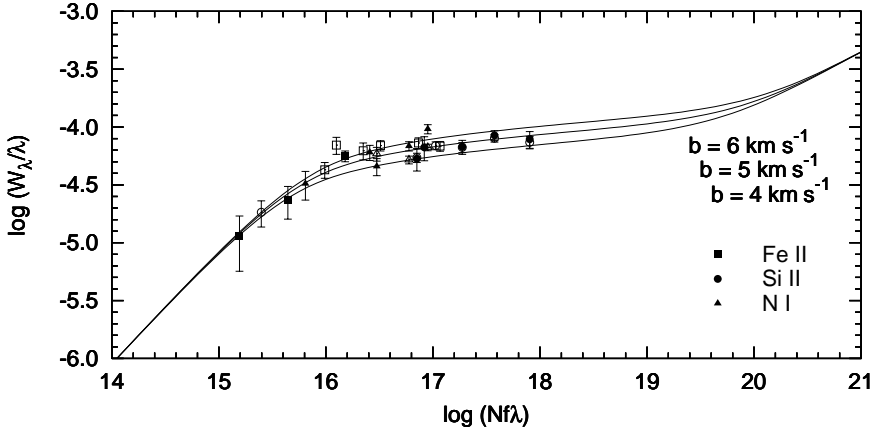}}
\hfill
\parbox[b]{55mm}{
\caption[]{The curve of growth for absorption by metals on the line of sight to \object{BD~+39\,3226} (component A) is shown. Plotted are only \ion{Si}{ii}, \ion{Fe}{ii}, and \ion{N}{i} which actually define the curve of growth. Filled symbols represent {\sl ORFEUS} data, open symbols those from the {\sl IUE}. The drawn curves represent single cloud absorption with indicated $b$-value }
\label{curgro1}
}
\end{figure*}

Table \ref{eqw1} lists the measured equivalent widths for component A.
The lines of Fe\,{\sc ii}, Si\,{\sc ii}, and N\,{\sc i} were used to determine the $b$-value of 5$\pm$1~km\,s$^{-1}$ of the curve of growth for metals.
 Then column densities of other ions were obtained by fitting their equivalent width to the curve.
Fig. \ref{curgro1}  shows the curve of growth, Table \ref{colden1} gives the resulting column densities.
 The errors take the uncertainties in the $b$-value and in the individual equivalent widths into account.
 For ions with data points only in the flat part of the curve of growth the column densities have large uncertainties.

\subsection{Neutral hydrogen}

\begin{figure*}[t!]
\resizebox{12cm}{!}{\includegraphics{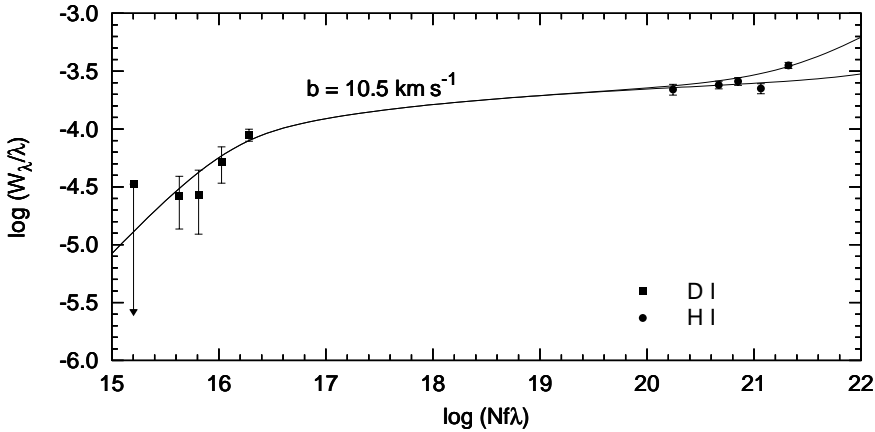}}
\hfill
\parbox[b]{55mm}{
\caption[]{The curve of growth for H\,{\sc i} and D\,{\sc i} towards \object{BD~+39\,3226}. In the damping part the curves split, the lower curve is for Ly~$\kappa$, the upper one is for Ly~$\delta$. Note that the D\,{\sc i} points lie essentially on the Doppler part of the curve of growth }
\label{curgro2}}
\end{figure*}

\begin{figure}[t]
 \resizebox{\hsize}{!}{\includegraphics{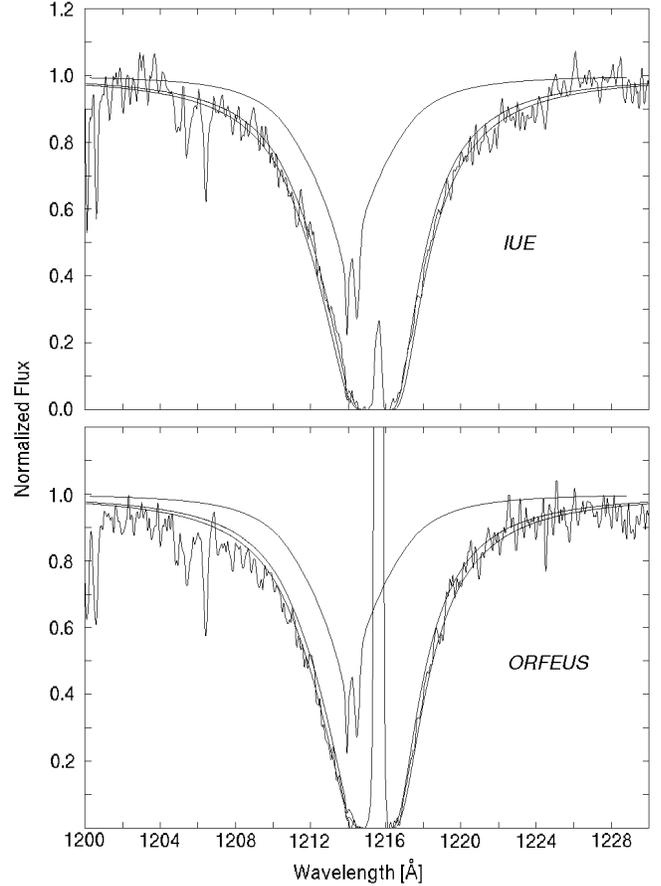}}
 \caption{Theoretical fits to the Lyman $\alpha$ line in the {\sl IUE} and the {\sl ORFEUS} spectrum. In both spectra the residual intensity near the centre of the absorption has been set to zero. The contribution of the stellar He\,{\sc ii} line  located $\simeq1.57$\,\AA~towards shorter wavelengths as calculated from an atmospheric model is included in the fits and plotted seperately. The two shown fits represent $\log N$(H\,{\sc i})$=20.0$ and $20.1$. Near the centre the geocoronal emission peak is visible  }
 \label{lyalph}
\end{figure}

\begin{figure}[t!]
 \resizebox{\hsize}{!}{\includegraphics{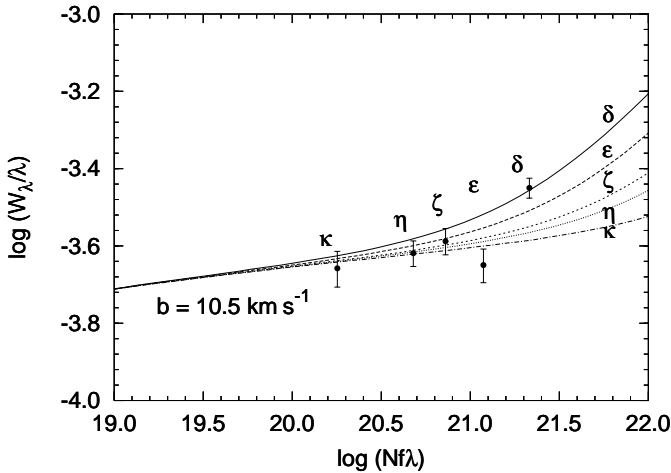}}
 \caption{Curve of growth for \ion{H}{i} in detail. The data points and the corresponding theoretical curves are labeled}
 \label{curgro3}
\end{figure}  

\begin{figure}[t!]
 \resizebox{\hsize}{!}{\includegraphics{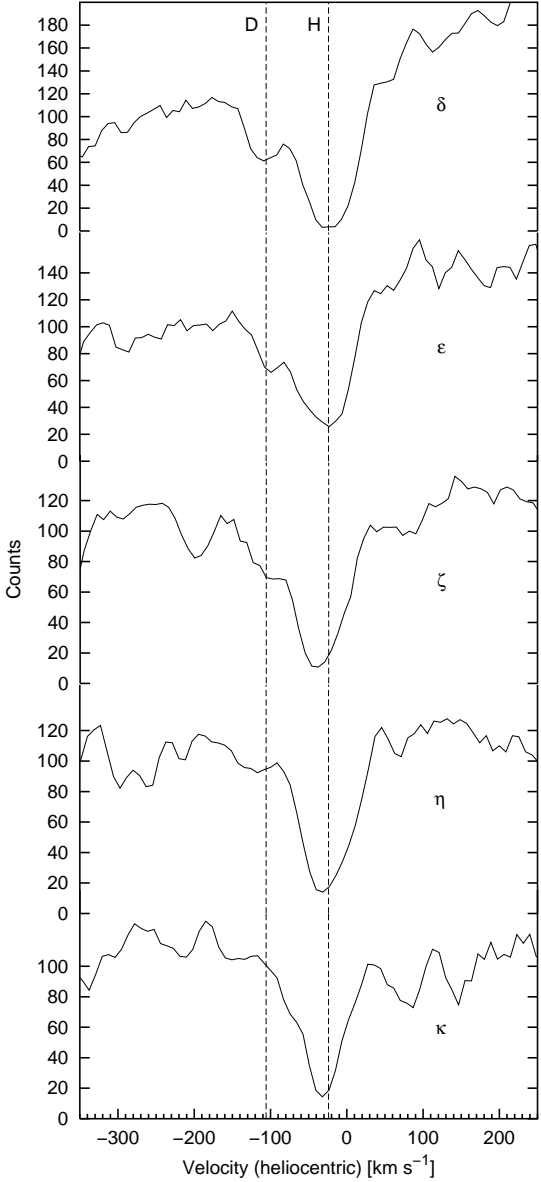}}
 \caption{The Lyman lines used in our curve of growth analysis in the {\sl ORFEUS} spectrum. The positions of D and H absorption by component A ($v_{\rm rad}\simeq-24$) are marked}
 \label{lycomp}
\end{figure}

 We determined the H\,{\sc i} column density in two different ways.

 First we compared theoretical Voigt profiles convolved with a gaussian
instrumental profile to the Ly ${\alpha}$ line in the {\sl ORFEUS} and the {\sl IUE} spectrum.  
 This line is always fully damped, therefore the $b$-value is unimportant and in case of a single velocity component only the column density remains as a parameter. Even small changes in the column density have a clear effect on the profile. We estimate the accuracy in $\log N$ as $\pm0.1$. 
 Problems arise because of the weaker component B  at $\Delta\lambda{\simeq}-0.21$\,{\AA}  and stellar He\,{\sc ii} absorption at  $\Delta\lambda{\simeq}-1.57$\,{\AA}  which both are unresolved.  Component B should have only a weak influence on the column density, probably $\leq 0.04$ dex. 
  The stellar line was calculated from an atmospheric model ($T_{\rm eff}=45000$~K, $\log g=5.5$, $n$(He)$=99$\%, $n$(H)$=1$\%) and included in the fit.
 The uncertainty in $N$(H\,{\sc i}) due to the stellar model is small, because significant errors in the strength of the calculated stellar line would have made the fit profile asymmetric with regard to the measured profile.
 An additional background substraction was applied in both spectra of the Lyman $\alpha$ line to correct for some residual intensity ($\approx5.5$\% in the {\sl ORFEUS} spectrum) near the centre.    
 The result is plotted in Fig. \ref{lyalph}. 
It is only possible to give a total column density (for component A and B), which is $N_{\rm H} = 1.12^{+0.29}_{-0.23}{\cdot}10^{20}$\,cm$^{-2}$.

\begin{table}[t!]
\caption[]{Hydrogen and deuterium equivalent widths. Wavelengths and oscillator strengths are taken from Morton (\cite{mor}). \ion{H}{i} and \ion{D}{i} Ly $\epsilon$ have been corrected for the influence of \ion{Fe}{ii} absorption at 937.652\,{\AA}, \ion{H}{i} and \ion{D}{i} Ly $\zeta$ for the influence of the H$_2$ We Q(1), 4-0 line at 930.574\,{\AA} }  
\label{eqw2} 
\begin{tabular}{lllll}
\hline\noalign{\smallskip}
Ion & Lyman line & Wavelength [\AA] & $\log f{\lambda}$ & $W_{\lambda}$ [m{\AA}]  \\ \noalign{\smallskip}\hline\noalign{\smallskip}
\ion{H}{i} & $\delta$ & $949.7431$ & $1.122$ & $337.0\pm20.1$ \\
           & $\epsilon$ & $937.8035$ & $0.864$ & $218.0\pm21.0$\\
           & $\zeta$  & $930.7483$ & $0.651$ & $240.3\pm18.9$ \\
           & $\eta$ & $926.2257$ & $0.470$ & $222.8\pm17.0$\\  
           & $\kappa$ & $919.3514$ & $0.043$ & $202.1\pm21.5$\\
\ion{D}{i} & $\delta$ & $949.4847$ & $1.122$ & \hspace{0.15cm}$84.6\pm10.0$ \\
           & $\epsilon$ & $937.5484$ & $0.865$ & \hspace{0.15cm}$47.2\pm17.5$ \\
           & $\zeta$ & $930.4951$ & $0.651$ & \hspace{0.15cm}$24.8\pm16.1$ \\
           & $\eta$ & $925.9737$ & $0.470$ & \hspace{0.15cm}$24.4\pm11.7$\\
           & $\kappa$ & $919.101\,^*$ & $0.043$ & $<30.7$\\
\noalign{\smallskip}\hline\noalign{\smallskip}  
\end{tabular}
\noindent
$^*$~Wavelength calculated from \ion{H}{i} Ly $\kappa$.
\end{table}

 To confirm this value, we also applied the curve-of-growth analysis to the Lyman series from Ly\,${\delta}$ to Ly\,${\kappa}$, except for Ly\,${\theta}$ and Ly\,${\iota}$ which seem distorted.
 The Ly\,$\beta$ and Ly\,$\gamma$ lines have strong damping wings which, together with the further structure of the spectrum, do not allow the determination of reliable equivalent widths.  
 Higher Lyman lines are also visible but may be blended with stellar He\,{\sc ii}, because the distance between these lines is smaller than the stellar radial velocity. Besides, near the Lyman edge it is not possible to set the continuum properly.
 We measured the equivalent width of Ly\,${\delta}$ by comparing the line with computed two-component-profiles (Voigt profiles convolved with the instrumental Gauss profile), one component for  the H\,{\sc i}  and one for the D\,{\sc i} line.   
 For the other lines the damping wings are neglegible and we used two-component Gauss fits.

\begin{figure}
 \resizebox{\hsize}{!}{\includegraphics{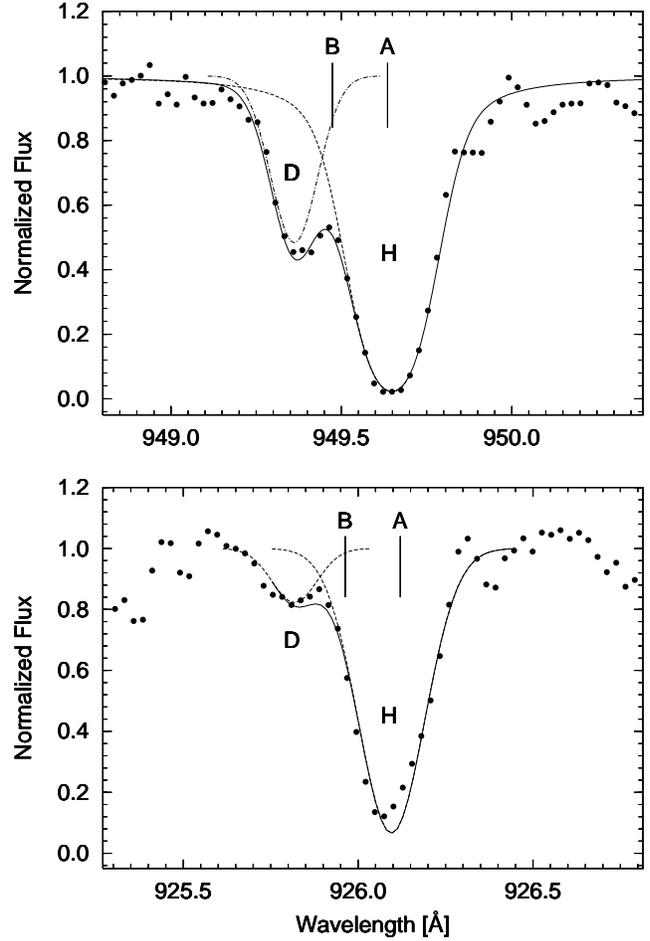}}
 \caption{For two of the Lyman lines a 500 km\,s$^{-1}$ wide section of the spectrum is displayed. {\it Top}: Deuterium and hydrogen Lyman $\delta$ line in the {\sl ORFEUS} spectrum (data points) with a two-component fit (Voigt profile convolved with instrumental Gauss). 
{\it Bottom}: Deuterium and hydrogen Lyman $\eta$ line (data points) with a two-component Gauss profile.  
 In both plots the positions of the velocity components A and B found in the metal lines are marked for the \ion{H}{i} lines. Component B has only neglegible effect on the absorption profile }
 \label{lydel}
\end{figure}

  We note that for the higher Lyman series lines ($\delta$-$\kappa$) the instrumental profile degrades the true absorption such that residual light is expected near the bottom of the profiles (see Fig. \ref{lycomp} and \ref{lydel}).
 A calculation shows this to be at the level of up to $10$\%.
 In addition, also in these lines geocoronal emission is present but not readily recognizable in the profiles considered
 (in Lyman $\alpha$ and $\beta$ it is clearly present).
 Since the absolute level of the contamination is not reliably known we have refrained from corrections.
 
 In case of Ly $\epsilon$ the \ion{Fe}{ii} line situated between the \ion{H}{i} and the \ion{D}{i} line at $937.652$ {\AA} was modeled and used as a third, fixed component in the fit. 
 An analogous procedure was necessary for the H$_2$ Werner Q(1), 4-0 line lying at $\lambda = 930.574$\,{\AA} between the \ion{H}{i} and \ion{D}{i} Lyman $\zeta$ lines. 
 Attempts to include velocity component B by additional fit components showed that it has neglegible influence on the line shape.  
 Examples for the fits are shown in Fig.\,{\ref{lydel}}.   
 Though the velocity structure is not resolved in the Lyman series, a single-cloud curve of growth should be sufficient since there is one clearly dominant cloud. 
 As expected the H\,{\sc i} curve of growth has a significantly higher Doppler-velocity than the metals' curve due to the much smaller atomic mass of hydrogen.  For a given $b$ each measured equivalent width and its error correspond to a column density with an error depending on the slope of the curve of growth. 
   We calculated $\log N_{\rm H}$ for different $b$-values from the weighted mean of the 5 column densities resulting from the 5 measured equivalent widths.  The least mean square deviation is found for $b = 10.5\pm1$~km\,s$^{-1}$, leading to $N_{\rm H} = 1.6^{+0.9}_{-0.6}{\cdot}10^{20}$\,cm$^{-2}$.     
 Fig.\,\ref{curgro2} and \ref{curgro3} show the curve for H\,{\sc i}.

 The Lyman {$\alpha$} fit and the curve of growth analysis give consistent results, so a mean value of $N_{\rm H} = 1.20^{+0.28}_{-0.22}{\cdot}10^{20}$\,cm$^{-2}$ can be derived. 

\subsection{Deuterium}

 The absorption of deuterium was investigated in 5 lines (see Table \ref{eqw2}). The \ion{D}{i} Lyman $\delta$ and $\eta$ lines are shown in Fig. \ref{lydel}.
 Along with the H\,{\sc i} data, the D\,{\sc i} equivalent widths are plotted in Fig. \ref{curgro2}. 
 For \ion{D}{i} Ly ${\kappa}$ only an upper limit can be given which is rather high due to the continuum uncertainty.  
 The Doppler parameter is only of minor importance for the determination of the D\,{\sc i} column density because the data points lie mainly on the linear part of the curve of growth.
 The deuterium data points seem to suggest a somewhat higher $b$-value than $10.5$\,km\,s$^{-1}$ but we used the same curve as for hydrogen since we expect $b_{\,\rm D} \leq b_{\,\rm H}$.    
  The weighted mean of the column densities derived for $b=10.5$~km\,s$^{-1}$ from the four measured $W_{\lambda}$ values
  is $N_{\rm D} = 1.45^{+0.50}_{-0.38}{\cdot}10^{15}$\,cm$^{-2}$.

\subsection{Molecular hydrogen} 

 The {\sl ORFEUS} spectra also contain a large number of absorption lines
from molecular hydrogen. 
  Only one component is visible here.
 The average radial velocity of $18$ lines measured in the echelle orders $50$-$59$ is
$v_{\rm{helio}}=-27.1\pm{0.9}$~km\,s$^{-1}$, slightly different from the velocity found for the metal absorption lines.
 The metal radial velocity of $v_{\rm{helio}}=-23.9\pm{1.0}$~km\,s$^{-1}$, was measured as the average value of $20$ lines in the echelle orders $42$-$60$.
 Since we have not found any obvious systematic velocity shift between different orders, a possible explanation may be the presence of an additional weak unresolved absorption component in the metal lines.

We have determined equivalent widths using trapezium fits.
No better quality of the results would be achievable from fitting gaussian 
profiles since most of the lines are only weak and do not show clear profiles.
The equivalent width for rotational states $J\ge5$ are similar to the strength 
of noise peaks, so only upper limits are determined.
Results are presented in Table \ref{H2_ew}.

\begin{figure*}[ht!]
  \resizebox{12cm}{!}{\includegraphics{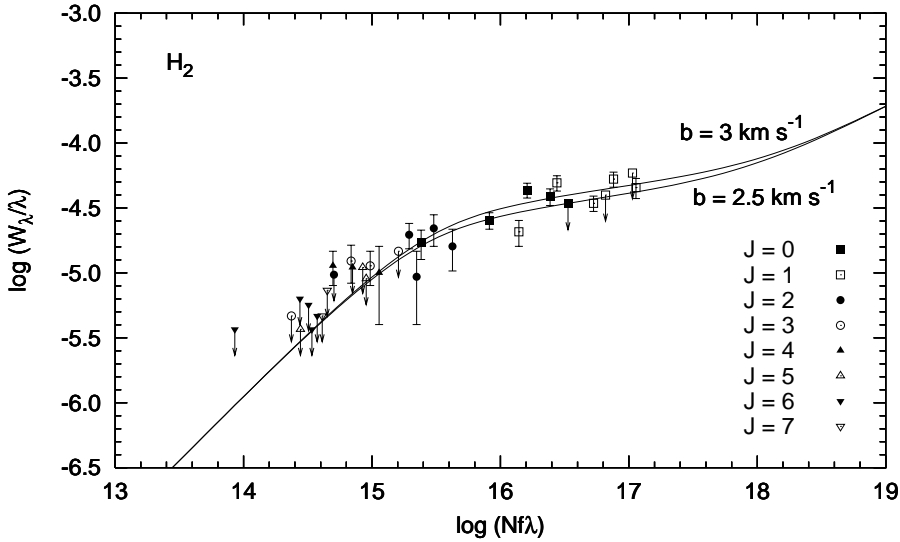}}
  \hfill
  \parbox[b]{55mm}{
    \caption[]{Curve of growth for the different excitation levels of H$_2$.
               Downward arrows denote upper limits for the equivalent width.
               The $b$-value is based on the absorption in levels $J=0$ to $2$}
    \label{CoG_H2}}
\end{figure*}

Column densities for the different rotational excitation levels are derived 
from these equivalent widths by fitting to a theoretical curve of growth 
(Fig. \ref{CoG_H2}).
The $b$-value is restricted by the lower $J$ levels to $2.5$ to 3~km\,s$^{-1}$.
The column densities for $J=0$ and $1$ lie in the flat part of the curve of 
growth and therefore are sensitive to variations of the $b$-value,
enlarging the errors for their column densities.
The higher rotational levels are located on the doppler part of the curve
which in principle allows a good definition of column densities. 
However, most of these lines are weak and have larger errors in the
equivalent widths which again leads to larger errors in the column densities.
For $J\ge5$ only upper limits can be given.
The resulting column densities are listed in Table \ref{H2_ew}.

The population of the lower rotational states of molecular hydrogen 
is determined by collisional excitation, following a Boltzmann distribution.
The excitation temperature can therefore be derived as
\begin{equation}
T_{mn}=\frac{E_n-E_m}{2.303\,k}
       \frac{1}{\log\left(N_m/g_m\right)-\log\left(N_n/g_n\right)},
\end{equation}
where the statistical weight $g_J$ is equal to $(2J+1)$, multiplied by a
factor of 3 for $J$ odd in regard to the triplet nature of ortho-H$_2$.
For the upper levels UV photons have to be considered as the primary source of
excitation (Spitzer \& Zweibel \cite{spitzer_zweibel}; Spitzer et al. 
\cite{spitzer_cochran}).
An equivalent excitation temperature can be derived here, but this has by
no means the physical implications of an actual kinetic temperature.

We have determined excitation temperatures by using an error weighted least
square fit. 
A temperature of $T_{02}=130\pm1$~K is derived for the lower levels $J=0$ to 
$2$, for the upper levels we find an upper limit in the equivalent excitation 
temperature of $T_{37}<1800$~K.

\begin{figure}[ht!]
  \resizebox{\hsize}{!}{\includegraphics{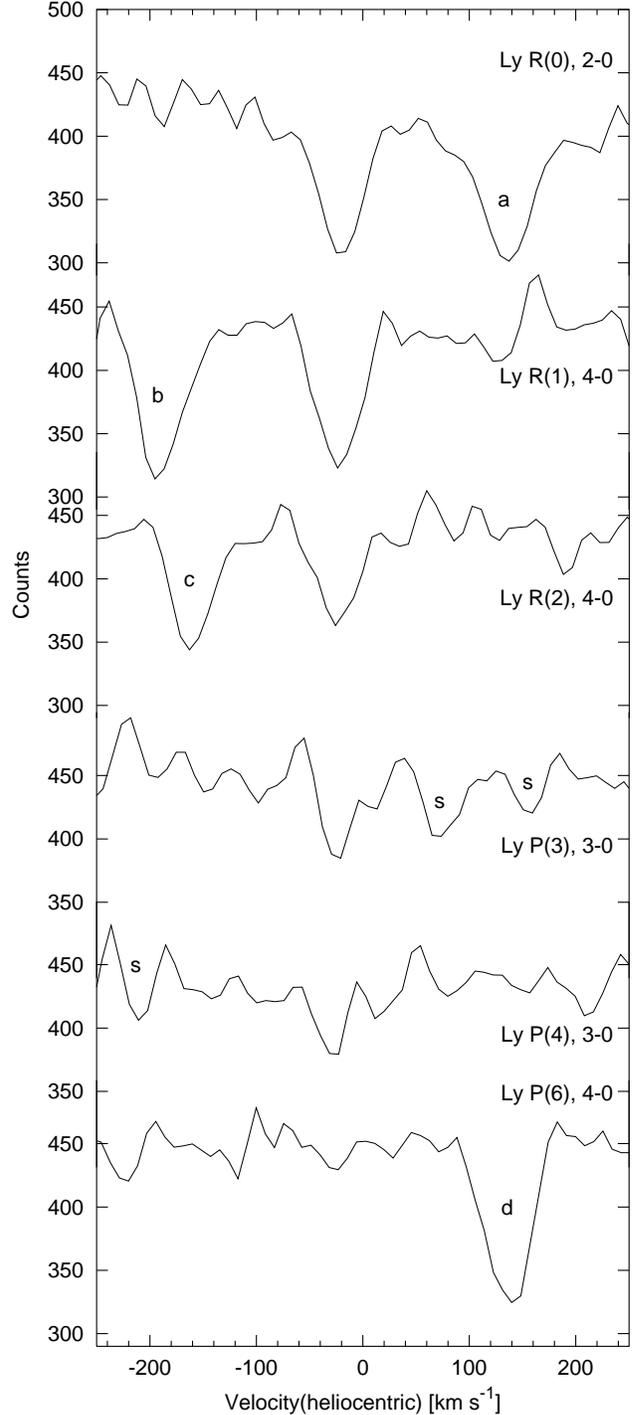}}
  \caption[]{Overview of the spectra for some H$_2$ transitions.
             The spectra shown are filtered by a wavelet algorithm.
             The H$_2$ absorption at 
             $v_{\rm{helio}}\simeq-27$~km\,s$^{-1}$ 
             is clearly visible for the transitions with $J\le4$. 
             At higher rotational states the equivalent widths are similar to 
             the strength of noise peaks.
             Additional absorption lines marked are: 
             a) Ly R(1), 2-0; b) Ly R(0), 4-0; c) Ly P(1), 4-0; 
             s) features of probably stellar origin, which could not be 
             unambigiously identified
            }
  \label{Spectra_H2}
\end{figure}

\begin{figure}[t]
\resizebox{\hsize}{!}{\includegraphics{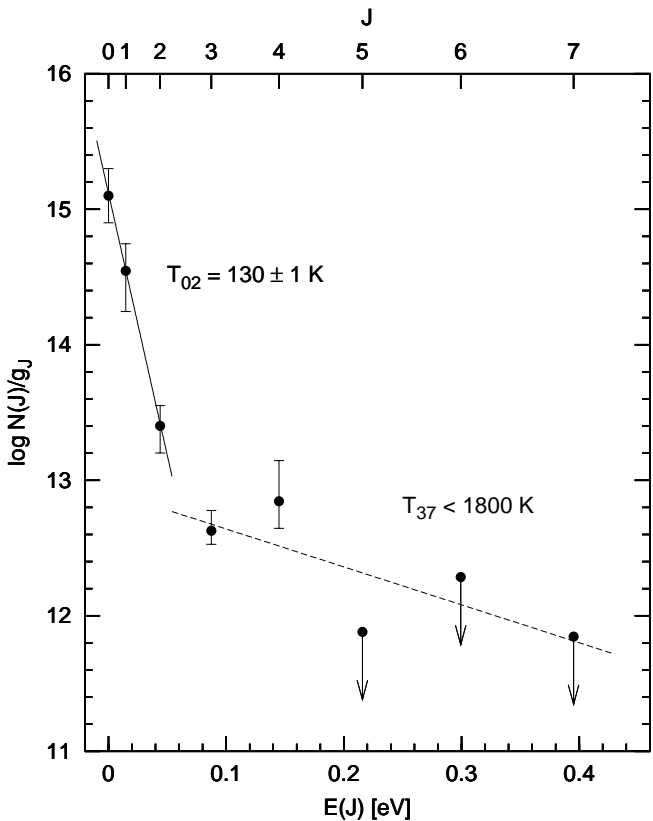}}
\caption[]{Excitation plot for the H$_2$ column densities.
           Plotted is the statistically weighted column density vs. 
           excitation energy of the rotational states.
           A clear separation of collisionally excited levels ($J\le2$) and of
           UV photon excited levels ($J\ge3$) is visible. 
           The lines denote least square fits to an assumed Boltzmann 
           distribution for levels $m$ to $n$ with the appropriate temperature
           $T_{mn}$.
           The error given for the Boltzmann excitation temperature $T_{02}$
           is the standard deviation of the fit. 
           For $T_{37}$ only an upper limit can be given
          }
\label{T_ex}
\end{figure}

Looking at Fig. \ref{T_ex}, where the column density $N_J$, weighted by the
statistical weight $g_J$, is plotted against the excitation energy 
of the rotational state $J$, a clear distinction is visible between the 
collisionally Boltzmann excited levels $J\le2$ and the UV photon excited levels
$J\ge3$.
Although the upper levels scatter around the fit, a much smaller slope is 
obvious, compared with the lower levels.

The H$_2$ ortho-to-para ratio (OPR) derived from the by far most prominent rotational states 
$J=0$ and $1$ is $N(1)/N(0)\simeq2.5$. 
The excitation temperature calculated using the lowest para-states $J=0$ and 
$2$ is $T_{02}=130$~K. 
This value essentially coincides with the temperature determined from the OPR 
of $T_{01}=134$~K. 
We therefore can assume that we see thermalized, purely collisionally excited
gas in the excitation levels up to $J=2$, following the Boltzmann 
distribution  (see Dalgarno et al. \cite{dalg}).

For the higher excitation levels information is less clear due to the
scatter in the data points.
Fitting the ortho- and para-states seperately we get $T_{35}<870$~K and 
$T_{46}<1400$~K, so the excitation temperature for the upper states probably 
can be restricted to $<1400$~K rather than $<1800$~K as stated above.

\section{Abundances}
\subsection{Metals, dust, and $H_2$}

\begin{table}[ht!]
\begin{minipage}[t]{9cm}
\begin{flushleft}
\caption[]{H$_2$ equivalent widths and column densities $N_J$ [cm$^{-2}$]}  
           
\label{H2_ew}
\begin{tabular}{rrlrc}
\hline\noalign{\smallskip}
\multicolumn{1}{c}{Transition} & \multicolumn{1}{c}{$\lambda$ $\left[\mbox{\AA}\right]$} & \multicolumn{1}{c}{$f^a$} & $W_{\lambda}$ $\left[\mbox{m\AA}\right]$ & Notes$^b$ \\
\noalign{\smallskip}
\hline
\noalign{\smallskip}
\multicolumn{5}{l}{$J=0$ $(g_J=1, E_J=0)$; $\log N_J=15.10\pm0.20$} \\
\noalign{\smallskip}
\hline\noalign{\smallskip}
Ly\hspace{1.15mm}R(0), 0-0 & $1108.128$ & $0.00173$  & $18.6\pm 4.8$ &   \\
Ly\hspace{1.15mm}R(0), 1-0 & $1092.194$ & $0.00596$  & $27.7\pm 4.0$ &   \\
Ly\hspace{1.15mm}R(0), 2-0 & $1077.138$ & $0.0119$   & $46.7\pm 6.1$ &   \\
Ly\hspace{1.15mm}R(0), 3-0 & $1062.883$ & $0.0182$   & $41.1\pm 6.0$ &   \\
Ly\hspace{1.15mm}R(0), 8-0 & $1001.826$ & $0.0266$   & $      <34.0$ & b \\
\noalign{\smallskip}
\hline\noalign{\smallskip}
\multicolumn{5}{l}{$J=1$ $(g_J=9, E_J=0.01469)$; $\log N_J=15.50^{+0.20}_{-0.30}$} \\
\noalign{\smallskip}
\hline\noalign{\smallskip}
We Q(1), 0-0               & $1009.772$ & $0.0238$   & $53.2\pm 7.1$ &   \\
Ly\hspace{1.15mm}R(1), 1-0 & $1092.732$ & $0.00403$  & $22.6\pm 5.1$ &   \\
We Q(1), 1-0               & $ 986.798$ & $0.0364$   & $44.7\pm 7.9$ &   \\
Ly\hspace{1.15mm}R(1), 2-0 & $1077.698$ & $0.00809$  & $53.2\pm 7.2$ &   \\
We Q(1), 2-0               & $ 966.097$ & $0.0349$   & $      <56.8$ & b \\
Ly\hspace{1.15mm}P\hspace{0.35mm}(1), 3-0 & $1064.606$ & $0.00584$  & $15.6\pm 5.0$ &   \\
Ly\hspace{1.15mm}R(1), 4-0 & $1049.958$ & $0.0160$   & $36.1\pm 4.9$ &   \\
Ly\hspace{1.15mm}P\hspace{0.35mm}(1), 7-0 & $1013.434$ & $0.0205$   & $      <40.2$ & b \\
\noalign{\smallskip}
\hline\noalign{\smallskip}
\multicolumn{5}{l}{$J=2$ $(g_J=5, E_J=0.04394)$; $\log N_J=14.10^{+0.15}_{-0.20}$} \\
\noalign{\smallskip}
\hline\noalign{\smallskip}
We Q(2), 0-0               & $1010.941$ & $0.0238$   & $22.3\pm 6.1$ &   \\
Ly\hspace{1.15mm}R(2), 1-0 & $1094.244$ & $0.00367$  & $      <10.6$ &   \\
We Q(2), 2-0               & $ 967.278$ & $0.0348$   & $15.5\pm 5.5$ & n \\
Ly\hspace{1.15mm}R(2), 4-0 & $1051.497$ & $0.0147$   & $20.7\pm 4.6$ &   \\
Ly\hspace{1.15mm}R(2), 5-0 & $1038.690$ & $0.0170$   & $ 9.7\pm 3.8$ &   \\
\noalign{\smallskip}
\hline\noalign{\smallskip}
\multicolumn{5}{l}{$J=3$ $(g_J=21, E_J=0.08747)$; $\log N_J=13.95^{+0.15}_{-0.10}$} \\
\noalign{\smallskip}
\hline\noalign{\smallskip}
Ly\hspace{1.15mm}P\hspace{0.35mm}(3), 1-0 & $1099.788$ & $0.00240$  & $       <5.2$ & n \\
Ly\hspace{1.15mm}P\hspace{0.35mm}(3), 3-0 & $1070.142$ & $0.00721$  & $13.2\pm 4.3$ &   \\
We Q(3), 4-0               & $ 933.581$ & $0.0193$   & $      <13.7$ &   \\
Ly\hspace{1.15mm}P\hspace{0.35mm}(3), 5-0 & $1043.498$ & $0.0104$   & $11.8\pm 3.5$ &   \\
\noalign{\smallskip}
\hline\noalign{\smallskip}
\multicolumn{5}{l}{$J=4$ $(g_J=9, E_J=0.14491)$; $\log N_J=13.80^{+0.30}_{-0.20}$} \\
\noalign{\smallskip}
\hline\noalign{\smallskip}
Ly\hspace{1.15mm}P\hspace{0.35mm}(4), 3-0 & $1074.313$ & $0.00733$  & $12.2\pm 3.6$ &   \\
Ly Q(4), 4-0               & $ 935.959$ & $0.0192$   & $ 9.4\pm 5.6$ &   \\
Ly\hspace{1.15mm}P\hspace{0.35mm}(4), 6-0 & $1035.184$ & $0.0108$   & $      <11.4$ &   \\
\noalign{\smallskip}
\hline\noalign{\smallskip}
\multicolumn{5}{l}{$J=5$ $(g_J=33, E_J=0.21575)$; $\log N_J<13.40$} \\
\noalign{\smallskip}
\hline\noalign{\smallskip}
We Q(5), 1-0               & $ 994.924$ & $0.0361$   & $       <9.0$ &   \\
We Q(5), 2-0               & $ 974.286$ & $0.0346$   & $      <10.7$ &   \\
Ly\hspace{1.15mm}P\hspace{0.35mm}(5), 5-0 & $1052.499$ & $0.0105$   & $       <3.9$ &   \\
\noalign{\smallskip}
\hline\noalign{\smallskip}
\multicolumn{5}{l}{$J=6$ $(g_J=13, E_J=0.29940)$; $\log N_J<13.40$} \\
\noalign{\smallskip}
\hline\noalign{\smallskip}
Ly\hspace{1.15mm}R(6), 1-0 & $1109.859$ & $0.00307$  & $       <4.1$ &   \\
Ly\hspace{1.15mm}P\hspace{0.35mm}(6), 4-0 & $1071.497$ & $0.0102$   & $       <6.8$ &   \\
Ly\hspace{1.15mm}R(6), 6-0 & $1041.711$ & $0.0143$   & $       <4.8$ &   \\
Ly\hspace{1.15mm}P\hspace{0.35mm}(6), 6-0 & $1045.805$ & $0.0122$   & $       <5.9$ &   \\
Ly\hspace{1.15mm}R(6), 8-0 & $1019.101$ & $0.0133$   & $       <3.7$ &   \\
\noalign{\smallskip}
\hline\noalign{\smallskip}
\multicolumn{5}{l}{$J=7$ $(g_J=45, E_J=0.39521)$; $\log N_J<13.50$} \\
\noalign{\smallskip}
\hline\noalign{\smallskip}
Ly\hspace{1.15mm}R(7), 5-0 & $1059.992$ & $0.0134$   & $       <7.7$ &   \\
Ly\hspace{1.15mm}P\hspace{0.35mm}(7), 6-0 & $1052.386$ & $0.0123$   & $       <4.9$ & \\
\noalign{\smallskip}
\hline\noalign{\smallskip}
\multicolumn{5}{l}{$^a$ $f$ values from Morton \& Dinerstein (\cite{mor:din})} \\
\multicolumn{5}{l}{$^b$ b: line is possibly blended; n: spectrum has locally small S/N} \\
\end{tabular}
\end{flushleft}
\end{minipage}
\end{table}

The metal abundances in component A are given in Table \ref{colden1}. Since only a total hydrogen column density for components A and B could be measured, the abundances may be too low by $\leq 0.04$ dex.
 In presence of the relatively large uncertainties this effect seems neglegible.  
 
 As can be seen in Table \ref{colden1} the depletion factors $\log\,d = \log(N/N_{\rm H}) - \log(N/N_{\rm H})_{\sun}$ of the elements vary between $-0.3$ and $-1.7$. 
 This moderate depletion indicates a low abundance of dust along the line of 
sight.

The total amount of H$_2$ seen on this line of sight is 
$N(\mbox{H}_2)=4.8^{+2.0}_{-1.6}{\cdot}10^{15}$\,cm$^{-2}$, which gives an abundance relative 
to H\,{\sc i} of $-4.40^{+0.17}_{-0.20}$\,dex. 
With this low value the total amount of hydrogen,
$N(\mbox{H\,{\sc i}}+\mbox{H}_2)
=[N(\mbox{H\,{\sc i}})+2N(\mbox{H}_2)]
=1.2{\cdot}10^{20}$\,cm$^{-2}$,
is hardly larger than the value given above for just H\,{\sc i}.

 The extinction towards BD~+39~3226 is given by Dworetsky et al. (\cite{dwor})
as $E(B-V)=0.05$, which again indicates a low dust abundance.
 In addition, the fractional abundance of molecular hydrogen 
$f=2N(\mbox{H}_2)/\left[N(\mbox{H\,{\sc i}})+2N(\mbox{H}_2)\right]$ 
has a low value of $\log f=-4.08$.
 The gas to color excess ratio for our line of sight of 
$N\left(\mbox{H {\sc i}}+\mbox{H}_2\right)/E(B-V)=2.3\cdot10^{21}$~cm$^{-2}$\,mag$^{-1}$ is slightly lower than the typical value for the galactic intercloud
gas of $5.0\cdot10^{21}$~cm$^{-2}$\,mag$^{-1}$ given by 
Bohlin et al. (\cite{bohlin78}).
 This deviation, though, can be explained by uncertainties in the $E(B-V)$ 
value and should not be overinterpretated. 
 We can conclude that most of the absorption seen on this line of sight arises
from intercloud gas with only little contribution from the clouds having
$T\simeq130$~K.

 Savage et al. (\cite{savage77}) have generated a plot of $\log f$ vs. $E(B-V)$
for their survey of $70$ stars. 
 For lines of sight with $E(B-V)<0.08$, they find only very small fractions of H$_2$ down to values $\log f$ of $-6$. 
 This is in line with the tight connection between the existence of dust grains
and of molecules forming on their surfaces. 
 Our low value fits well into this group of stars of low fractional abundance.

In their plot, Savage et al. (\cite{savage77}) show theoretical curves from
equilibrium models of Black for different hydrogen densities, radiation 
densities, and molecule formation rates. 
According to this, our value is described by Black's ``Model $2$'', which 
assumes a density $n(\mbox{H\,{\sc i}}+\mbox{H}_2)$ of 100~cm$^{-3}$ at 
a temperature of about $100$~K, in accordance to the temperature derived above.
The radiation density at $1000$~\AA\ has then a value of 
$4\cdot10^{-17}$~erg\,cm$^{-3}$\,\AA$^{-1}$ if the molecule formation rate is
$10^{-17}$~cm$^3$\,s$^{-1}$.

\subsection{Deuterium}
 From the \ion{H}{i} and \ion{D}{i} column densities we obtain a deuterium abundance on this line of sight of $D/H = 1.2^{+0.5}_{-0.4} {\cdot} 10^{-5}$.        
 We also tried to identify absorption lines of HD, but without success.   
 Taking the H$_2$ column density of only $4.8{\cdot}10^{15}$\,cm$^{-2}$ one would expect HD column densities as low as about $10^{11}$\,cm$^{-2}$, which is definitely below the detection limit of $\simeq 10^{13}$\,cm$^{-2}$ for the spectra available.
 It can be ruled out that any significant amount of deuterium is hidden in HD.
 Deuterium depletion by interactions with dust grains is unlikely in a warm ISM component with low gas-to-dust ratio.  
  
 The abundance is a value for the whole line of sight, possibly an average over more than one cloud with different $D/H$ ratios. The uncertainties are rather large, 
 but the value is reliable within its limitations. There was no need for assumptions about the velocity structure of the absorption (as it is necessary for line modelling) since the hydrogen column density determination is mainly based on fully damped absorption while the deuterium lines lie in the linear part of the curve of growth. In both cases the curve of growth is insensitive to effects arising from clouds of different $b$-values and column densities.
 Furthermore, a model of the stellar spectrum was used only for Lyman $\alpha$ fitting. 
 The uncertainty arising from modelling is neglegible because the centres of the stellar and interstellar absorption lines are well separated due to the high radial velocity of the star.

   Varying scattered light may affect the measured equivalent widths, as mentioned in Sect. 2.  
  However, we do not find evidence for systematic background substraction errors beyond the $5$\% level in all analyzed H and D lines and the scatter of the equivalent widths is not larger than expected for the estimated errors.
 The effect of systematic errors would be rather small.
 For example, if all H and D equivalent widths were larger by $10$~\% than measured
 due to wrong background substraction, the D\,{\sc i} column density also would rise by $\approx10$\% while the H\,{\sc i} column density resulting from curve of growth analysis and Lyman $\alpha$ fitting would change even less.
 Thus the $D/H$ ratio would be increased by less than $10$~\% and would remain well within the given range.

\section{Concluding remarks}

 McCullough (\cite{mccu}) finds all {\sl Copernicus} and {\sl IUE}\, ISM deuterium abundance measurements to be consistent with $D/H = (1.5\pm0.2){\cdot}10^{-5}$. 
 This value was confirmed for the local interstellar cloud by {\sl GHRS} observations towards \object{$\alpha$ Aur} (Linsky et al. \cite{lins}) and \object{HR 1099} (Piskunov et al. \cite{pisk}) yielding $D/H = 1.60^{+0.14}_{-0.19}{\cdot}10^{-5}$ and $1.46\pm0.09{\cdot}10^{-5}$ respectively. 
  For a summary of {\sl GHRS} measurements see Linsky (\cite{lins2}). 
 McCullough's list of included data contains $6$ measurements towards nearby stars ($d\,{\leq}\,14$\,pc) observed only in Ly\,$\alpha$ and $8$ {\sl Copernicus} observations of each $2$ to $5$ Lyman lines towards more distant stars ($d\,{\geq}\,79$\,pc).    
 The mean of the results of the latter observations is $D/H\simeq1.4{\cdot}10^{-5}$. 
 Our result based on 5 deuterium lines lies below this value though it is consistent within its uncertainty.
            
 Recent determinations of $D/H$ for example by Vidal-Madjar et al. (\cite{vidmad}) towards \object{G191-B2B} using the {\sl GHRS}  ($D/H =1.12_{-0.08}^{+0.08}{\cdot}10^{-5}$), G\"olz et al. (\cite{golz}) in a preliminary analysis of {\sl ORFEUS II} observations of \object{BD +28\,4211} ($D/H = 0.8^{+0.7}_{-0.4}{\cdot}10^{-5}$) or Jenkins et al. (\cite{jenk2}) using {\sl IMAPS} spectra of \object{$\delta$ Ori} ($D/H = 0.74^{+0.19}_{-0.13}{\cdot}10^{-5}$) indicate smaller values towards early type stars with $d \geq 69$\,pc. As Vidal-Madjar et al. and Jenkins et al. state, a possible explanation may be blending of the Local Interstellar Cloud (LIC) contribution with that of more distant deuterium-poor clouds. 
Small scale spatial inhomogeneities of $D/H$ could be the result of the production of significant amounts of deuterium in stellar flares, as Mullan \& Linsky (\cite{mul:lin}) suggest. 
  This discussion remains open since Sahu et al. (\cite{sahu}) found $D/H=1.60^{+0.39}_{-0.27}\cdot10^{-5}$ towards \object{G191-B2B} from a {\sl STIS} spectrum, a value consistent with a homogeneous $D/H$ in the LISM.  
 
 However this may be, the various determinations of $D/H$ all are consistent with $D/H\approx(1.2\pm0.3){\cdot}10^{-5}$.
 The uncertainty in each individual value is influenced by the spectral resolution of the instrument, by the number of D lines used, and by the actual D column density.
 Our result from the new instrument {\sl ORFEUS} using $5$ lines at $30$\,km\,s$^{-1}$ resolution adds to these determinations .

\acknowledgements
PR is supported by grant 50 QV 9701 3 of DLR (formerly DARA), OM is supported by grant Bo 779/24 of DFG. {\sl ORFEUS} data analysis in Bamberg is supported by the DLR under grant 50 QV 97026.
 We thank the T\"ubingen and Heidelberg {\sl ORFEUS} team for their support and basic data reduction.

\end{document}